\documentclass[iop]{emulateapj}
\usepackage{graphicx,subfigure,amsmath, amsfonts, amssymb,aas_macros,footnote,color,epstopdf,nnfootnote,times}
\usepackage[bookmarks=false]{hyperref}
\newcommand{\ha}{H$\alpha$}

\newcommand{\msun}{M$_{\odot}$}

\newcommand{\atlas}{ATLAS$^{\rm 3D}$}
\shorttitle{NGC~1266 as a Quenched SF System}
\shortauthors{K. Alatalo et al.}

\begin{document}

%%%%%%%%%%%%%%%%%%%%%%%%%%%%%%%%%%%%%%%%%%%%%%%
% TITLE AND AUTHOR

\title{NGC~1266 as a Local Candidate for Rapid Cessation of Star Formation}

\author{K. Alatalo$^{1,2}$,
 K. Nyland$^{3}$, 
 G. Graves$^{1,4}$, 
 S. Deustua$^{5}$, 
 K. Shapiro Griffin$^{6}$, 
 P.-A. Duc$^{7}$, 
 M. Cappellari$^{8}$, 
 R.~M. McDermid$^{9,10,11}$, 
 T.~A. Davis$^{12}$, 
 A.~F. Crocker$^{13}$,
 L.~M. Young$^{3,14}$, 
 P. Chang$^{15}$, 
 N. Scott$^{16,17}$, 
 S.~L. Cales$^{18}$, 
 %start of alphabetical
 E. Bayet$^{8}$,
 L. Blitz$^{1}$, 
 M. Bois$^{19}$, 
 F. Bournaud$^{7}$, 
 M. Bureau$^{8}$, 
 R.~L. Davies$^{8}$, 
 P.~T. de Zeeuw$^{12,20}$, 
 E. Emsellem$^{12,21}$, 
 S. Khochfar$^{22}$, 
 D. Krajnovi\'c$^{23}$, 
 H. Kuntschner$^{8}$, 
 R. Morganti$^{24,25}$, 
 T. Naab$^{26}$, 
 T. Oosterloo$^{24,25}$, 
 M. Sarzi$^{27}$, 
 P. Serra$^{24,28}$, 
 A. Weijmans$^{29}$
 }

%%%%%%%%%%%%%%%%%%%%%%%%%%%%%%%%%%%%%%%%%%%%%%%
% ABSTRACT

\begin{abstract}
We present new Spectrographic Areal Unit for Research on Optical Nebulae ({\tt SAURON}) integral-field spectroscopy and {\em Swift} Ultraviolet Optical Telescope (UVOT) observations of molecular outflow host galaxy NGC~1266 that indicate NGC~1266 has experienced a rapid cessation of star formation.  Both the {\tt SAURON} maps of stellar population age and the {\em Swift} UVOT observations demonstrate the presence of young ($< 1$ Gyr) stellar populations within the central 1 kpc, while existing Combined Array for Research in Millimeter-wave Astronomy (CARMA) CO(1--0) maps indicate that the sites of current star formation are constrained to the inner few hundred parsecs of the galaxy only.  The optical spectrum of NGC~1266 from \citet{moustakas+06} reveal a characteristic post-starburst (K+A) stellar population and Davis et al. (2012) confirm that ionized gas emission in the system originate from a shock.  Galaxies with K+A spectra and shock-like ionized gas line ratios may comprise an important, overlooked segment of the post-starburst population, containing exactly those objects in which the AGN is actively expelling the star-forming material.  While AGN activity is not the likely driver of the post-starburst event that occurred 500 Myr ago, the faint spiral structure seen in the Hubble Space Telescope (HST) Wide-field Camera 3 (WFC3) {\em Y-, J-} and {\em H-}band imaging seems to point to the possibility of gravitational torques being the culprit.  If the molecular gas were driven into the center at the same time as the larger scale galaxy disk underwent quenching, the AGN might be able to sustain the presence of molecular gas for $\gtrsim 1$ Gyr by cyclically injecting turbulent kinetic energy into the dense molecular gas via a radio jet, inhibiting star formation.
%Deep optical images of NGC~1266 lack evidence that NGC~1266 has undergone a major merger, but the presence of faint tidal streams leaves open the possibility of a minor merger in the recent past.  The minor merger hypothesis is also supported by faint spiral structure seen in Hubble Space Telescope Wide Field Camera 3 (WFC3) near-infrared imaging.  We hypothesize that NGC~1266 can be explained via an induced-quenched star formation cessation process.  First, a minor merger transports the molecular gas to the center.  Eventually, the gas triggers the AGN, which is then able to drive the gas out of the system.  This process may be able to explain the morphologically undisrupted population of post-starburst quasars.

%We present {\tt SAURON} and {\em Swift} UVOT observations of the local quiescent galaxy, NGC~1266.  These observations indicate that NGC~1266 has seen an outside-in cessation of star formation, with the sites of young stars being far more extended as compared to the molecular gas.  Integrated spectra of the NGC~1266 nucleus \citep{moustakas+06} revealed a characteristic K+A \citep{dressgunn86} poststarburst stellar population, despite the presence of ionized gas emission.  \citet{davis+12} showed that the ionized gas emission in the center of NGC~1266 is due to the shock associated with the outflow, indicating that galaxies with K+A spectra and shock-like ionized gas diagnostics may comprise an important, perhaps overlooked, segment of the poststarburst population.
\end{abstract}

%%%%%%%%%%%%%%%%%%%%%%%%%%%%%%%%%%%%%%%%%%%%%%%
% SUBJECT HEADINGS

%%%%%%%%%%%%%%%%%%%%%%%%%%%%%%%%%%%%%%%%%%%%%%%
% ACTUAL PAPER
\section{Introduction}
%%%%%%%%%%% KRISTINA'S NOTES %%%%%%%%%%%%
%I am worried about the poststarburst classification a bit . . . NGC~1266 is very luminous in the infrared.  Lots of refs on NED to support this.  NGC~1266 has a non-negligable SFR in other tracers, as you know and quoted in your 2011 paper.  NGC~1266 also has the spiral feature (very similar to NGC 3801 - see Hota et al. 2012 - this is an S0 galaxy with a molecular outflow and a faint spiral patter visible in GALEX data . . . sound familiar?).  Anyways, Hota et al. 2012 argue that the spiral feature has a merger origin (after kinematically modeling).  Could NGC~1266 not be as ``quiescent" as we thought?

% Add in size to table of VLBA stuff

The present-day galaxy population has a bimodal color distribution, with a genuine lack of intermediate-color galaxies \citep{strateva+01,baldry+04}.  The lack of galaxies in the ``green valley'' suggests that galaxies migrate rapidly between the \nnfoottext{\\$^{1}$Department of Astronomy, Hearst Field Annex, University of California - Berkeley, California 94720, USA\\
$^{2}$Infrared Processing and Analysis Center, California Institute of Technology, Pasadena, California 91125, USA\\
$^{3}$Physics Department, New Mexico Tech, Socorro, NM 87801, USA\\
$^{4}$Department of Astrophysical Sciences, Princeton University, Princeton, New Jersey 08544, USA\\
$^{5}$Space Telescope Science Institute, 3700 San Martin Drive, Baltimore, MD 21218, USA\\
$^{6}$Space Sciences Research Group, Northrop Grumman Aerospace Systems, Redondo Beach, CA 90278, USA\\
$^{7}$Laboratoire AIM Paris-Saclay, CEA/IRFU/SAp -- CNRS -- Universit\'e Paris Diderot, 91191 Gif-sur-Yvette Cedex, France\\
$^{8}$Sub-Dept. of Astrophysics, Dept. of Physics, University of Oxford, Denys Wilkinson Building, Keble Road, Oxford, OX1 3RH, UK\\
$^9$Australian Astronomical Observatory, PO Box 296, Epping, NSW 1710, Australia\\
$^{10}$Department of Physics and Astronomy, Macquarie University, NSW 2109, Australia\\
$^{11}$Gemini Observatory, Northern Operations Centre, 670 N. A`ohoku Place, Hilo, HI 96720, USA\\
$^{12}$European Southern Observatory, Karl-Schwarzschild-Str. 2, 85748 Garching, Germany\\
$^{13}$Department of Physics and Astronomy, University of Toledo, Toledo, OH 43606, USA\\
$^{14}$National Radio Astronomy Observatory, Socorro, NM 87801, USA\\
$^{15}$Department of Physics, University of Wisconsin - Milwaukee, Milwaukee, WI 53201, USA\\
$^{16}$Centre for Astrophysics \& Supercomputing, Swinburne University of Technology, PO Box 218, Hawthorn, VIC 3122, Australia\\
$^{17}$Sydney Institute for Astronomy (SIfA), School of Physics, The University of Sydney, NSW, 2006, Australia\\
$^{18}$Department of Astronomy, Faculty of Physical and Mathematical Sciences, Universidad de Concepci\'{o}n, Casilla 160-C, Concepci\'{o}n, Chile\\
$^{19}$Observatoire de Paris, LERMA and CNRS, 61 Av. de l'Observatoire, F-75014 Paris, France\\
$^{20}$Sterrewacht Leiden, Leiden University, Postbus 9513, 2300 RA Leiden, the Netherlands\\
$^{21}$Universit\'e Lyon 1, Observatoire de Lyon, Centre de Recherche Astrophysique de Lyon and Ecole Normale Sup\'erieure de Lyon, 9 avenue Charles Andr\'e, F-69230 Saint-Genis Laval, France\\
$^{22}$Max-Planck Institut f\"ur extraterrestrische Physik, PO Box 1312, D-85478 Garching, Germany\\
$^{23}$Leibniz-Institut f\"ur Astrophysik Potsdam (AIP), An der Sternwarte 16, D-14482 Potsdam, Germany\\
$^{24}$Netherlands Institute for Radio Astronomy (ASTRON), Postbus 2, 7990 AA Dwingeloo, The Netherlands\\
$^{25}$Kapteyn Astronomical Institute, University of Groningen, Postbus 800, 9700 AV Groningen, The Netherlands\\
$^{26}$Max-Planck-Institut f\"ur Astrophysik, Karl-Schwarzschild-Str. 1, 85741 Garching, Germany\\
$^{27}$Centre for Astrophysics Research, University of Hertfordshire, Hatfield, Herts AL1 9AB, UK\\
$^{28}$CSIRO Astronomy and Space Science, Australia Telescope National Facility, PO Box 76, Epping, NSW 1710, Australia\\
$^{29}$Dunlap Institute for Astronomy \& Astrophysics, University of Toronto, 50 St. George Street, Toronto, ON M5S 3H4, Canada}
blue cloud and red sequence, requiring a rapid quenching of star formation (SF; \citealt{bell+04,Faber+07}).  Recent simulations have suggested that active galactic nuclei (AGN) may be capable of shutting down SF by heating and driving out gas \citep{springel+05,Hopkins+05,croton+06,debuhr+12}.  While circumstantial evidence for the quenching of SF via AGN feedback exists \citep{schawinski+07}, direct evidence has been scarce.  There are promising candidates for AGN-driven SF quenching in quasar hosts at $z\sim2$ \citep{nesvadba+08, cano-diaz+12}, but only limited information can be obtained from such distant objects.  
In the more local universe, the low-redshift quasar host Markarian~231 has recently been shown to exhibit a massive molecular outflow \citep{feruglio+10,fischer+10,aalto+12a}.  However, due to the high current SF rate, it is difficult to distinguish between a starburst and an AGN-driven origin for the outflow in this system.  Other nearby galaxies with candidate AGN-driven molecular outflows such as M51 \citep{matsushita+07}, NGC~3801 \citep{das+05, hota+12} and NGC~1377 \citep{aalto+12b} have been similarly controversial, with compact starbursts or the effects of a recent major merger potentially dominating over any AGN feedback.  Thus, examples of gas-rich systems unaltered by recent major mergers or strong starbursts are needed to better understand the role of AGN feedback in the SF history of local galaxies.  

Post-starburst galaxies (i.e., K+A and E+A galaxies; \citealt{dressgunn86,zabludoff+96,quintero+04,goto2005}) may be an ideal demographic to study in the search for evidence of direct AGN-driven SF quenching.  These galaxies tend to have undergone a rapid ($< 1$ Gyr) cessation of SF (although a rapid onset of SF is also capable of forming post-starbursts; \citealt{falkenberg+09}) and commonly lie in the green valley of the galaxy color-magnitude distribution.  While some post-starburst galaxies show obvious signs of disruption and are likely the products of recent major mergers, the reason behind the abrupt halt of SF in others is less clear \citep{cales+11}.  Proposed methods of SF quenching in non-merging systems include strangulation and ram pressure stripping (as a gas-rich galaxy falls into a cluster environment and experiences tidal effects and winds, respectively; \citealt{evrard91,fujita98,bekki+02,boselli+06}), harassment (in which the gravitational potential of a galaxy is perturbed by its cluster neighbors; \citealt{icke85,mihos95,bekki98,moore+96}), tidal stripping (tidal disruption of cold gas due to nearby galaxies, like what is seen in Hickson Compact Groups; \citealt{verdes-montenegro+01}), morphological quenching (the tendency of the molecular gas within a bulge-dominated system to be too stable against gravitational collapse to efficiently form stars; \citealt{martig+09}) and AGN-driven feedback in the radiative (ionization, heating and radiation pressure; \citealt{ciotti+07}) or mechanical mode (nuclear winds and jets; \citealt{ciotti+10}).  Strangulation, ram pressure stripping and harassment require a dense cluster environment, tidal stripping requires a group environment and therefore cannot explain solitary post-starbursts, and morphological quenching should not generate the massive molecular outflows observed in some nearby galaxies.  Thus, isolated post-starburst galaxies with actively outflowing gas provide snapshots of a critical phase of a galaxy's journey from star-forming to the red sequence.

The discovery of the massive molecular outflow ($M_{\mathrm{outflow}}=2.4\times10^{7}$ M$_{\rm \sun}$, \.{\em M} = 13 M$_{\odot}$ yr$^{-1}$) in NGC~1266 was originally reported in \citet{alatalo+11}.  Recently, \citet{davis+12} also presented their work on integral-field spectroscopic observations of the ionized gas in the nucleus of NGC~1266 and showed that the outflow is truly multiphase, consisting of ionized and atomic gas, molecular gas, X-ray emitting plasma and radio emitting plasma.  Finally, Nyland et al. (2013) report on the discovery of a high brightness temperature VLBA point source within the nucleus of the galaxy, providing definitive evidence of the presence of an AGN in the system.  The distance to NGC~1266 is taken from \atlas, 29.9 Mpc, for which 1\arcsec = 145 pc.

Here we report on the stellar population of NGC~1266, particularly how it has changed with time, how that change might relate to post-starburst objects in general, and how we might find other objects undergoing similar events.  In \S\ref{obs}, we describe the observations and data reduction of Hubble Space Telescope Wide-field Camera 3 (HST WFC3), {\em Swift} Ultraviolet Optical Telescope (UVOT), {\tt SAURON} and the 2.3m Bok telescope at Kitt Peak.  In \S\ref{disc}, we show that NGC~1266 contains a post-starburst-like stellar population, and we discuss the implications of NGC~1266's re-classification as a post-starburst galaxy.  In \S\ref{conclu}, we summarize our results and suggest future directions for NGC~1266 studies.

%%%%%%%%%%%%%%%%% OBSERVATIONS %%%%%%%%%%%%%%%%%%%%%
\section{Observations and Data Reduction}
\label{obs}

\subsection{Hubble Space Telescope (HST)}
Visible and infrared images of NGC~1266 were obtained with the Hubble Wide Field Camera 3 (WFC3) instrument on the HST in December 2011.   Table \ref{tab:hst} lists the HST datasets identification, instrument, channel, filter and exposure time.  All images are full frame, and were processed with the standard reduction pipeline {\tt CALWF3}.  Cleaned images were coadded, registered and scaled to a common pixel scale of 0.13 arcsec/pixel with {\tt MULTIDRIZZLE}.   The resulting drizzled images were flux calibrated and appear in Figure \ref{fig:wfc3color}.

\begin{figure*}[t!]
\centering
\includegraphics[width=7in,clip,trim=0.4cm 0.1cm 1.3cm 2.3cm]{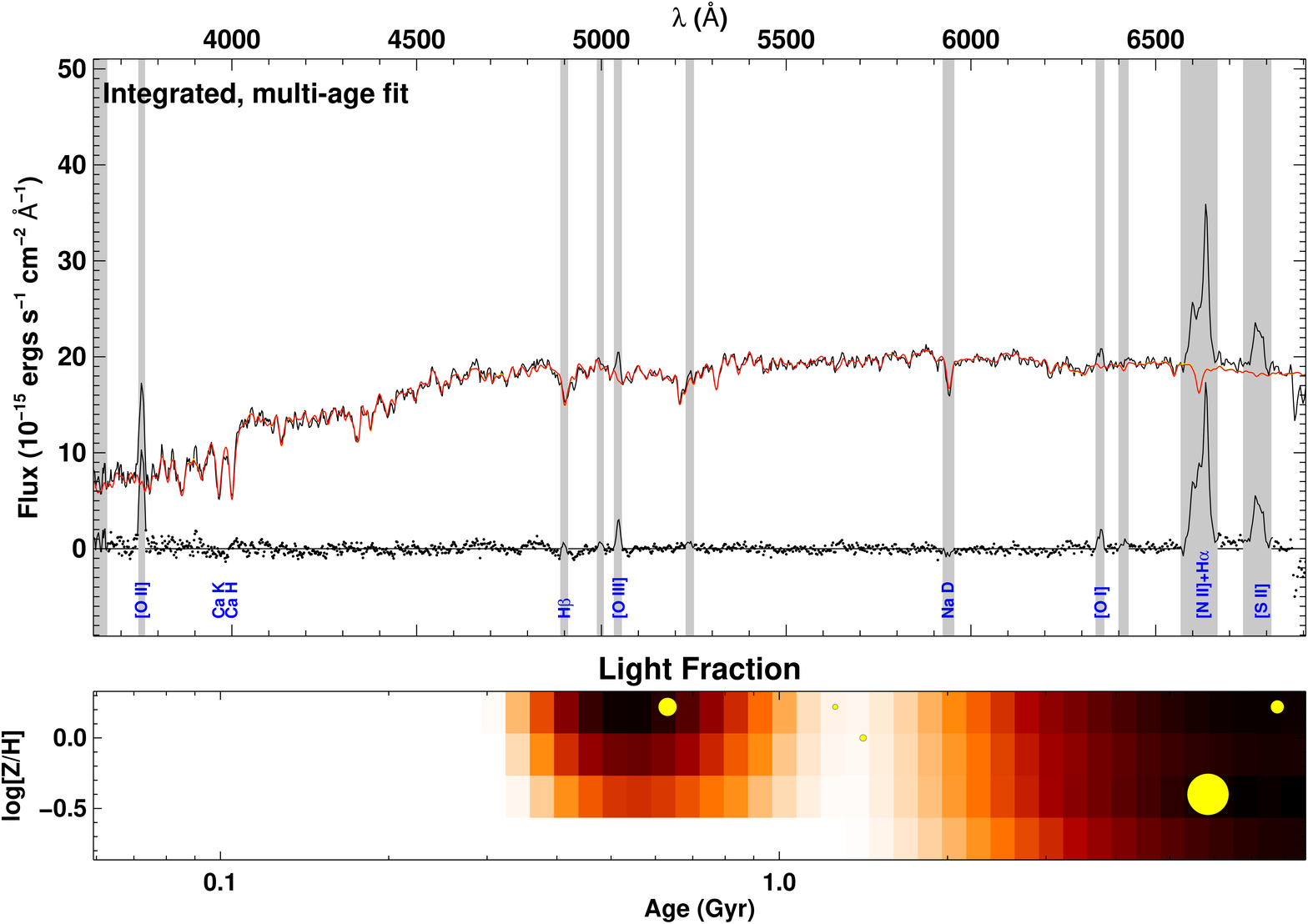}
\caption{\small {\bf (Top):} Integrated spectra (black) from 3800--6000 \AA\ are overlaid with the stellar model fits from MIUSCAT multiplied by a weighting supplied by {\tt pPXF} for both bursty (yellow) and continuous (red) modes, using a universal Kroupa Initial Mass Function \citep{kroupa01}.  Residuals (black points) and masked portions of the spectra (shaded gray) are also included, and well-known lines are identified in blue.  The parameters of the integrated fit did not vary between different libraries (MILES or MIUSCAT; \citealt{vazdekis+10}), though did depend on the chosen IMF (see Table \ref{tab:lightfrac}).  {\bf (Bottom):} The relative age-metallicity weights derived from a {\tt pPXF} fit to a continuous SF history using a universal Kroupa IMF (controlled by the line regularization weighting given through the {\tt REGUL} keyword in {\tt pPXF}; \citealt{capp+em04}) for the spectrum.  The redscale represents a linear light fraction weighting, with darker regions tending to have higher weights.  The results of a bursty SF history light fraction weights are shown as yellow points (of which there are 5), with their size representing the relative contribution of each age and metallicity to the fit.  Light fraction was derived using the $V$-band mass-to-light ratio for each metallicity and age bin using the universal Kroupa IMF.}
\label{fig:int_spec}
\end{figure*}

\begin{figure*}[t!]
\centering
\includegraphics[width=7in,clip,trim=0.4cm 0cm 1.3cm 2.3cm]{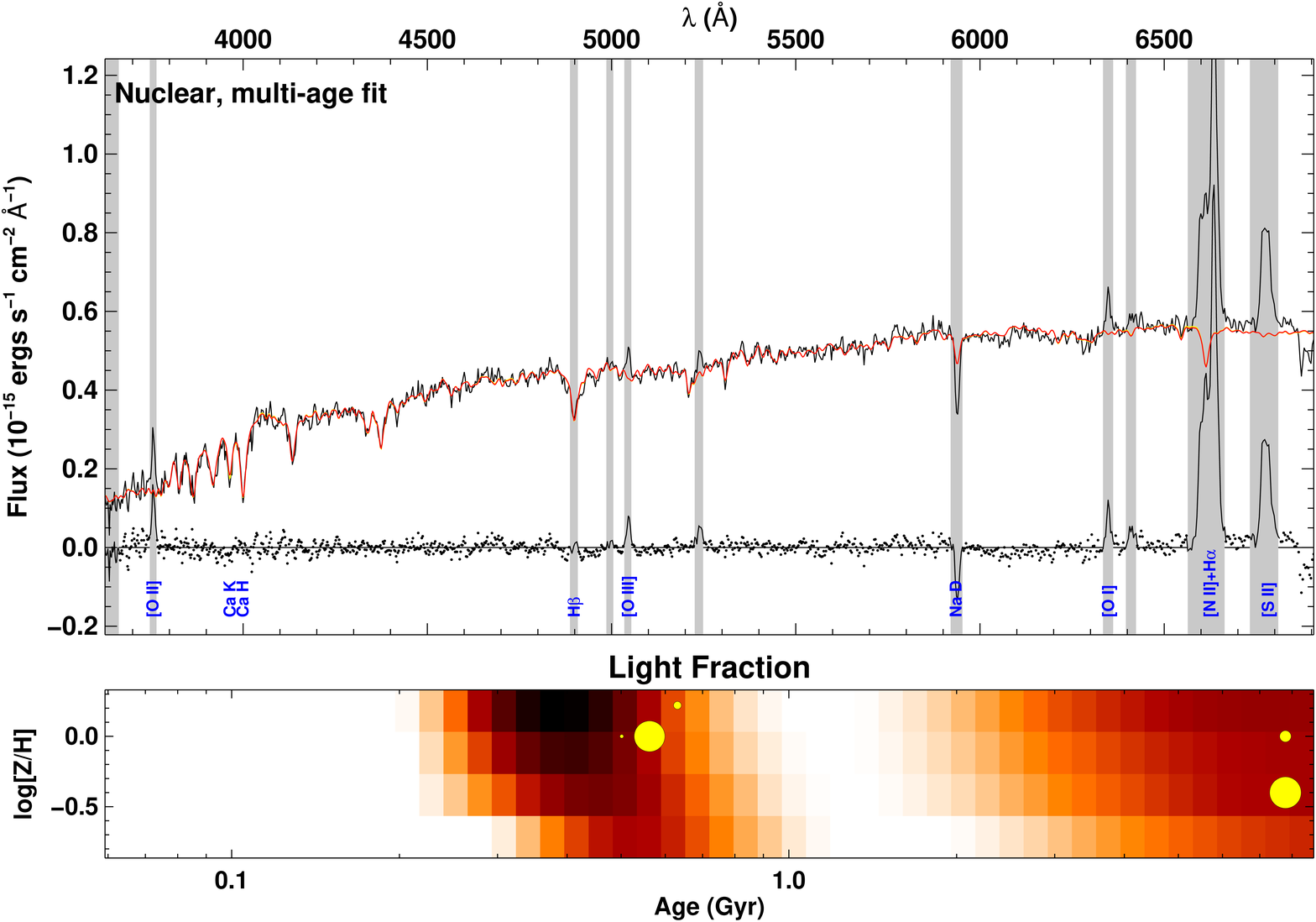}
\caption{\small {\bf (Top):} Nuclear spectra (black) from 3800--6000 \AA\ are overlaid with the stellar model fits from MIUSCAT multiplied by a weighting supplied by {\tt pPXF} for both bursty (yellow) and continuous (red) modes, using a universal Kroupa Initial Mass Function \citep{kroupa01}.  Residuals (black points) and masked portions of the spectra (shaded gray) are also included, and well-known lines are identified in blue.  The parameters of the nuclear fit did not vary between different libraries (MILES or MIUSCAT; \citealt{vazdekis+10}), though did depend on the chosen IMF (see Table \ref{tab:lightfrac}).  {\bf (Bottom):} The relative age-metallicity weights derived from a {\tt pPXF} fit to a continuous SF history using a universal Kroupa IMF (controlled by the line regularization weighting given through the {\tt REGUL} keyword in {\tt pPXF}; \citealt{capp+em04}) for the spectrum.  The redscale represents a linear light fraction weighting, with darker regions tending to have higher weights.  The results of a bursty SF history light fraction weights are shown as yellow points (of which there are 6), with their size representing the relative contribution of each age and metallicity to the fit.  Light fraction was derived using the $V$-band mass-to-light ratio for each metallicity and age bin using the universal Kroupa IMF.}
\label{fig:nuc_spec}
\end{figure*}

\begin{table}[h!]
\caption{HST WFC3 Observations (Program 12525, 3 orbits)}
\centering
\begin{tabular*}{7.3cm}{l c l l r}
\hline \hline
Obs ID & Instrument & Channel & Filter & Exp. Time \\ 
& & & & (seconds) \\
\hline
ibr702c6q & WFC3 & IR & F160W & 449 \\
ibr702c7q & WFC3 & IR & F140W & 449 \\
ibr702c9q & WFC3 & IR & F110W & 399 \\
ibr702cbq & WFC3 & IR & F160W & 449 \\
ibr702cdq & WFC3 & IR & F140W & 449 \\
ibr702cfq & WFC3 & IR & F110W & 399 \\
\hline \hline
\label{tab:hst}
\end{tabular*}
\end{table}

\subsection{Swift Ultraviolet Optical Telescope (UVOT)}
NGC~1266 was obverved with the {\em Swift} Ultraviolet-Optical Telescope (UVOT; \citealt{uvot}) under Target-of-Opportunity Program \#31376 on March 15 and 17, 2009 for 9840s and 14220s in the near-UV ($\lambda2600$) and far-UV($\lambda1928$) bands, respectively.  Data were reduced using the automatic pipeline and coadded using the {\em Swift} task {\tt UVOTIMSUM} from the HEASOFT package\footnote{HEASOFT software can be found at: http://heasarc.gsfc.nasa.gov/docs/software/lheasoft/} \citep{breeveld+10}.  The resultant images were then sky-subtracted, and the sky root mean square noise was determined using the IDL DAOPHOT task {\tt SKY}.  All discussions hereafter focus on the far-UV ($\lambda1928$) data.

\subsection{{\tt SAURON}}

The Spectrographic Areal Unit for Research on Optical Nebulae ({\tt SAURON}) is an integral-field spectrograph at the William Herschel Telescope (WHT; \citealt{bacon+01}). The {\tt SAURON} data of NGC~1266 were taken on the nights of 2008 January 10 and 11, as part of the \atlas\ observing campaign \citep{cappellari+11}. {\tt SAURON} covers the wavelength range 4810 - 5350 \AA\ with a spectral resolution of 105 km s$^{-1}$.  The SAURON observations were reduced using the standard \atlas\ pipeline \citep{cappellari+11} and the processed data cubes were Voronoi binned \citep{cappellari+03}.

In order to measure the absorption line strengths, the data were first processed using GANDALF \citep{sarzi+06}, providing the best-fit combination of absorption and emission lines.  The emission contribution was then subtracted from the {\tt SAURON} spectrum.  All details of the H$\beta$ absorption (H$\beta_{\rm abs}$) and single stellar population (SSP) modeling will be presented in McDermid et al. (2013), in prep.  To apply astrometry to the {\tt SAURON} images, the peak of the integrated {\tt SAURON} map was matched to the peak of the V-band image from the Spitzer Infrared Nearby Galaxy Survey (SINGS) \citep{kennicutt+03}.  Since the {\tt SAURON} wavelength range is nearest to V-band, this provided a reasonable match for the astrometry, allowing for WCS info based on the match to the V-band data to be written into the {\tt SAURON} H$\beta_{\rm abs}$ and stellar age headers.%  The astrometry was then applied to the {\tt SAURON} H$\beta_{\rm abs}$ and stellar age.%{\bf Also include the addition of WCS info to SAURON by comparing peak of V-band to peak of SAURON integrated}.

\subsection{Long-slit Spectroscopy}
We also provide a new analysis of long-slit spectroscopy originally published in \citet{moustakas+06}.  These data were originally obtained at the 2.3m Bok telescope on Kitt Peak using the Boller \& Chivens spectrograph, providing for spectral coverage between 3600 and 6900 \AA\ with 2.75 \AA\ pixels and a full-width at half maximum resolution of 8 \AA, through a $2\farcs5$ wide by $3\farcm3$ long slit.  A spectroscopic drift scan technique (scanning the slit across the galaxy while integrating) was used to construct an integrated spectrum. The drift scan length perpendicular to the slit for NGC~1266 was 55$''$, with a total exposure time of 2400 s.  The nuclear spectrum of NGC~1266 was obtained based on five-minute exposures using a fixed $2\farcs5\times2\farcs5$ slit aperture. The final flux calibrated spectra of the nucleus and integrated data for NGC~1266 were delivered in a FITS table to be analyzed for this paper, and further detail on the reduction and analysis techniques are available in \citet{moustakas+06}.  The spectral range of these observations allows for a more robust determination of stellar populations by providing a much larger set of Balmer absorption lines than the {\tt SAURON} observations, which only include H$\beta$.

%%%%%%%%%%%%%%%%% RESULTS %%%%%%%%%%%%%%%%%%%%%
\section{Methods \& Analysis}
\label{disc}

\subsection{The stellar composition of NGC~1266}
\label{spec_analysis}

%\begin{table}[b]
%\caption{NGC~1266 stellar population fitting results}
%\label{tab:ppxf}
%\begin{tabular*}{8.7cm}{ll l l l l l l l}
%\hline \hline
%\multicolumn{6}{l}{\bf Nuclear}\\
%\hline
%Age (Gyr) & 0.63 & 0.63 & 0.70 & 10.0 & 11.2 & 15.9\\
%Z (Z$_\odot$) & 1.0 & 1.7 & 1.7 & 1.7 & 1.7 & 1.7\\ 
%Weight & 0.23 & 0.13 & 0.14 & 0.04 & 0.14 & 0.12\\
%\hline
%\multicolumn{6}{l}{\bf Integrated}\\
%\hline
%Age (Gyr)  & 0.9 & 1.4 & 2.2 & 15.8 & 17.7 & 17.7\\
%Z (Z$_\odot$) & 1.0 & 1.0 & 0.4 & 1.7 & 0.4 & 1.7\\
%Weight & 0.11 & 0.44 & 0.08 & 0.22 & 0.05 & 0.06\\
%\hline \hline
%\vspace{-1mm}
%\end{tabular*}
%\small{The weighting results from a complete set of stellar age and metallicity libraries \citep{vazdekis+10} from fitting using {\tt pPXF}.  All templates were included in the fit, and the keyword {\tt REGUL} was used, in order to ensure the smoothest distribution for the fit.}
%\end{table}
%Nuclear: 41% 0.3 Gyr population, 69% 10 Gyr population 
%Drift55: 16% 0.3 Gyr, 84 % 10 Gyr

\begin{table*}[t!]
\begin{center}
\caption{NGC~1266 stellar population weights}
\label{tab:lightfrac}
\begin{tabular*}{15.5cm}{@{\extracolsep{\fill} }l c c c c c c}
\hline \hline
\multicolumn{5}{l}{\bf Total age distributions ({\tt pPXF}, MIUSCAT)}\\
\hline
& & Light& Mass & & Light & Mass\\
\multicolumn{2}{l}{\bf Kroupa Universal} & fraction & fraction & & fraction & fraction \\
\hline
Nuclear& $<2$ Gyr & 0.49 & 0.07 & $>10$ Gyr & 0.51 & 0.93\\
Integrated& $<2$ Gyr & 0.41 & 0.08 & $>10$ Gyr & 0.59 & 0.92\\
\hline
\multicolumn{5}{l}{\bf Kroupa Revised}\\
\hline
Nuclear& $<2$ Gyr & 0.54 & 0.09 & $>10$ Gyr & 0.46 & 0.91\\
Integrated& $<2$ Gyr & 0.40 & 0.07 & $>10$ Gyr & 0.60 & 0.93\\
\hline
\multicolumn{5}{l}{\bf Unimodal (Salpeter)}\\
\hline
Nuclear& $<2$ Gyr & 0.51 & 0.08 & $>10$ Gyr & 0.49 & 0.92\\
Integrated& $<2$ Gyr & 0.40 & 0.07 & $>10$ Gyr & 0.60 & 0.93\\
\hline
\multicolumn{5}{l}{\bf Bimodal}\\
\hline
Nuclear& $<2$ Gyr & 0.50 & 0.07 & $>10$ Gyr & 0.50 & 0.93\\
Integrated& $<2$ Gyr & 0.36 & 0.06 & $>10$ Gyr & 0.64 & 0.94\\
\hline
\multicolumn{5}{l}{\bf Total A and K-star distributions (K+A only)}\\
\hline
Nuclear& A (0.3 Gyr) & 0.85 & 0.12 & K (10 Gyr) & 0.15 & 0.88\\
Integrated& A (0.3 Gyr) & 0.68 & 0.12 & K (10 Gyr) & 0.32 & 0.88\\
\hline \hline
\vspace{-5.5mm}
\end{tabular*} \\
\end{center}
\small{The total normalized stellar contributions broken up into ``young'' ($<2$ Gyr) and ``old'' ($>10$ Gyr) age bins for the nuclear and integrated spectra for NGC~1266, with both mass fractions and luminosity fractions listed, fit to MIUSCAT templates assuming 4 different IMFs (listed above).  Although both spectra have young populations, it is clear that the stellar population in the nucleus has a larger fraction of young stars, as compared with the integrated spectrum.  The non-negligible fraction of young stars in NGC~1266 is independent of IMF choice.}  The results from a K+A fit are also included, showing that when restricted to 2 templates, NGC~1266 would be classified as a post-starburst system by \citet{quintero+04}.
\end{table*}

To determine the age composition of stars within NGC~1266, we used the spectra originally published in \citet{moustakas+06}.
We masked emission lines known to be part of the shock from \citet{davis+12} at [\ion{O}{2}]$\lambda3727$, [\ion{O}{3}]$\lambda5007$, H$\alpha$, H$\beta$, [\ion{N}{2}]$\lambda6583$ and [\ion{S}{2}]$\lambda6716$, as well as visible Na D absorption and a sky-line ($\lambda5577$).  We then used the Penalized Pixel-Fitting ({\tt pPXF}) IDL procedure \citep{capp+em04} to fit a set of stellar population templates from the MIUSCAT library \citep{vazdekis+12}, which is an update on the original MILES library \citep{falcon-barroso+11}, spanning \hbox{$-0.71 < [Z/H] < +0.20$} linearly in metallicity and logarithmically spanning 63 Myr to 17.8 Gyr logarithmically in age \citep{vazdekis+10}.  Four Initial Mass Functions (IMFs) were used to derive the differences in fractions across IMF choices, including Unimodal \citep{salpeter55}, Bimodal \citep{vazdekis+10}, and finally universal Kroupa and revised Kroupa \citep{kroupa01}.  The spectra fits were also done with {\tt pPXF} using the MILES library \citep{falcon-barroso+11}, in order to check the efficacy of the models and search for large diversions between model fits, with similarly selected [Z/H] and age parameters.  The MILES models derived equivalent mass fractions in each IMF as found by the MIUSCAT models.

In order to gauge the different possible star formation histories that could be present in NGC~1266, {\tt pPXF} was run with two different assumptions.  The first assumption is that NGC~1266 has had a monolithic star formation history, leading to a much smoother stellar population weight distribution, which was created using line regularization (setting the REGUL parameter in {\tt pPXF}; \citealt{capp+em04}).  Setting a nonzero REGUL keyword has the effect of pushing a star formation history toward a more linear shape in age and Z space, and effectively providing the smoothest fit among the many degenerate solutions that fit the data equally well.  This linearization is also able to allow us to explore the degeneracy amongst the various stellar population models.  The second was assuming a bursty star formation history (setting REGUL = 0), which the multiple young subcomponents of NGC~1266 seem to point to.  Both sets of assumptions create a bimodal stellar distribution, and both the bursty and smoothly varying star formation histories produced the same light fraction for young vs. old stars when fit with the MIUSCAT models.   Figures \ref{fig:int_spec} and \ref{fig:nuc_spec} show both the flux-calibrated long-slit spectra, the MIUSCAT universal Kroupa model fits as well as the light fraction weighting given to the component stellar ages and metallicities fitted by {\tt pPXF}.  These figures indicate that the distribution of models appears to be bimodal, with a ``young'' ($< 2$ Gyr) and ``old'' ($>$ 10 Gyr) stellar population.  As there is degeneracy in the $>10$ Gyr stellar population models, the weighting of stellar models are driven toward the edge of the distribution in NGC~1266, which has long been known to host (at least in part) a large mass of old stars.  Although we are unable to pinpoint exactly the metallicity and age of the ``older'' stellar population, {\tt pPXF} is able to provide a good estimate of the relative light contributions to the spectra.  Table \ref{tab:lightfrac} summarizes the results of fitting the integrated and nuclear spectra to the MIUSCAT models with four IMF choices.  The mass fractions were derived based on the output weights from the bursty star formation (REGUL=0) models, and light fractions were derived based on the mass-to-light ratios in $V$-band derived directly from the IMFs used.  Finally, a classical A/K fit was also run, limiting {\tt pPXF} to two templates, solar metallicity A (0.3 Gyr) and K (10.0 Gyr) templates, derived from the MIUSCAT Kroupa universal IMF set of stellar population models.

%We maximized temporal line regularization in our full template fit (similar to assuming a constant star formation rate (SFR) over time), as well as fit a bursty model (without line regularization).  In the constant SFR case, the models chosen by {\tt pPXF} tended to have a larger range of metallicities, with extra weight given to sub-solar values, which are somewhat degenerate with younger stellar templates.  Assuming that the SFR was bursty over time increased the total contribution of young stars, both in the nuclear and integrated spectra.  Figure \ref{fig:spec} gives the constant SFR {\tt pPXF} fit overlaid on the original spectrum.  It is clear from the fits that the distribution of stars is fairly bimodal, with both a contribution of young ($<2$ Gyr) and old ($>10$ Gyr) stars, with few signs of intermediate age stars (although this method has been shown to robustly recover intermediate age stars when present; McDermid et al, in preparation).  Nuclear emission is also present in the integrated spectrum, so it is possible that the fractional contribution of light from the young stars is concentrated in the nucleus.

%It is likely that many of the young stars present in the integrated spectrum are contributed from the nucleus.
%It is also clear that the nucleus of NGC~1266 has a larger contribution of young stars to its spectrum, and it is likely that the young stellar contribution present in the integrated spectrum are from the nucleus.

In order to run a classical K+A analysis, first described in \citet{dressgunn86}, and later in \citet{quintero+04}, we constrained the models available to {\tt pPXF} to A-star (0.3 Gyr) and K-star (10.0 Gyr) models with solar metallicity ([Z/H] = 0.0).  The inferred contribution of young stars to the spectrum is larger when we limit our models to K+A models (see Table \ref{tab:lightfrac}).  We find that the A-star mass contribution to all spectra is $\approx 10$\%, with 68--85\% of the light originating from the A-stars.  The classical K+A fitting seems to have slightly overestimated the total contribution of young stars within the NGC~1266 spectrum.  It is possible that constraining fits to these templates in other post-starburst searches might have overestimated the total contribution of young stars as well.
%.  The nucleus is more pronounced, with an A-star contribution of 24\%.  
%{\bf It is notable that the full stellar population fitting produced a larger mass fraction of $<2$ Gyr-aged stars than a simple K+A fit, by almost a factor of 2 in the nucleus and a factor of 4 in the outer spectrum.  

Using the MIUSCAT results from Figures \ref{fig:int_spec} and \ref{fig:nuc_spec}, the average age of the ``young'' stellar population is approximately 0.8 Gyr in the outer regions and $\approx 0.5$ Gyr in the nucleus, which is older than an A-star, but still much younger than the ``old'' stellar population.  It is possible that constraining fits to just these two templates is far too simplistic in understanding the star formation history (and possibly post-starburst phase) for galaxies.
%It is of note that the classical K+A fitting seems to have overestimated the total contribution of young stars within the NGC~1266 spectrum.  It is possible that constraining fits to these templates in other post-starburst searches might have overestimated the total contribution of young stars as well.
%In both the nuclear and integrated spectra, the Balmer absorption lines (H$\beta~\lambda4860$, H$\gamma~\lambda4340$ and H$\delta~\lambda4100$) are strong, indicating the presence of a young stellar population throughout the galaxy.  The Ca H and K lines, located just blue-ward of the 4000\AA\ break, are roughly equal in strength in the integrated spectrum (Fig. \ref{fig:spec}b).  However, the Ca H line is stronger than the K line in the nuclear spectrum (Fig. \ref{fig:spec}a) due to the contribution from H$\epsilon$ absorption (whose wavelength lies in the same position as Ca H). 
%The integrated spectra show a relatively young stellar population consisting of 1.0--2.5~Gyr old stars, with the intermediate age (2.5~Gyr) stars contributing over half of the light.  In contrast, the nuclear spectrum contains a significant (66\%) contribution from a young ($<1$ Gyr), post-starburst ``A-star'' population, and only 34\% contribution from older ($>10$ Gyr) stars. The distribution of ages quoted here should be viewed as an approximate guide to the fraction of old, intermediate age, and young stars, rather than an exact multi-burst star formation history.

Future work utilizing broadband HST imaging to create a high-resolution multicolor map of NGC~1266, in conjunction with SED modeling of stellar populations \citep{dacunha+08}, will be able to create a map of the star formation history within the galaxy, thus providing a much more detailed look at the way in which star formation has shut down throughout the disk of NGC~1266.  New, deep, high-resolution data will be capable of tracing star-forming regions in the molecular disk as well as the environment surrounding the central AGN.

\subsection{Spatial distribution of SF}
\label{sf_distribution}

%Given the highly compact distribution of molecular gas in NGC~1266 (A+11), it is interesting to examine the spatial distribution of the star formation activity and of the young stellar populations.  
Because of the compact nature of the molecular gas, we examine the comparative extent of both current and recent star formation to infer the progression of the star formation in NGC~1266. The molecular gas not involved in an outflow is concentrated in a central disk of radius $<250$ pc \citep{alatalo+11}, so any current or future star formation activity must be concentrated within this radius.

\begin{figure}[t]
\centering
\includegraphics[width=3.4in,clip, trim=0cm 1cm 0cm 1cm]{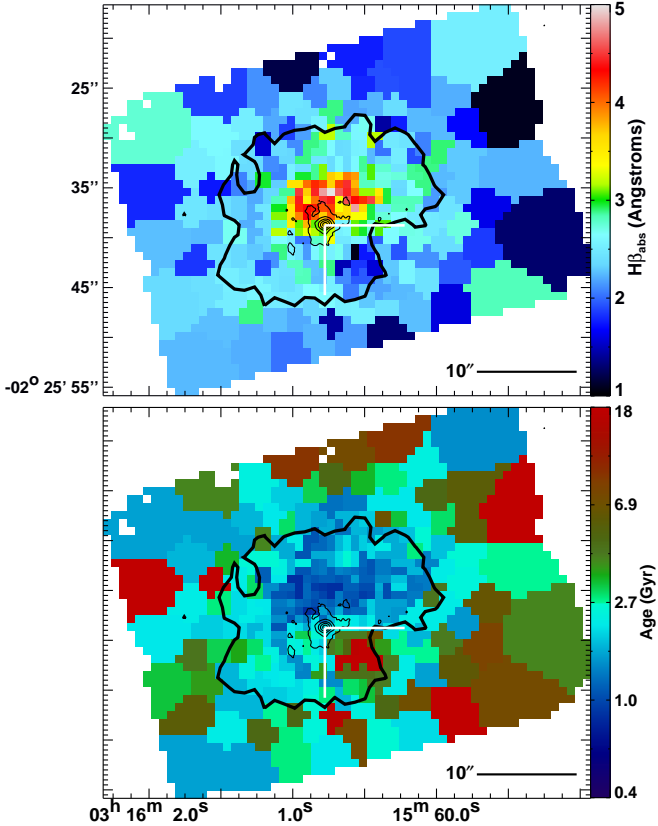}
\caption{{\bf (Top):} {\tt SAURON} map of the H$\beta$ absorption.  All absorption measurements plotted have EW $>1.4$ \AA\ , indicating young stars.  Overlaid are both the $3\sigma$ {\em UV} boundary (outermost contour) and the CO(1--0) image from Fig. \ref{fig:sf_distrib} (inner contours).  {\bf (Bottom):} {\tt SAURON} SSP model age of the stars within NGC~1266 (McDermid et al. 2014, in prep), in logarithmic scale.  It is clear from these data that the young stellar population within the galaxy is more extended than the sites of current star formation. The {\em UV} image, the H$\beta$ absorption map and the SSP-derived age map show a much larger region of young stars, closer to a few kpc, than the molecular gas, likely indicating that star formation has migrated inward.  It is of note that the H$\beta$ absorption along the axis of the outflow (approximately to the southeast of the CO emission) could be filled in by strong ionized gas emission (see \citealt{davis+12}).  White lines represent the approximate vertices of the outflow defined by the obscuration seen in the HST B-band image \citep{nyland+13}, and are places where strong shock emission are found in \citet{davis+12}.  The correspondence between the lack of H$\beta$ absorption and placement of the outflow vertex argues that these decrements are likely due to extinction, rather than an intrinsic asymmetry in the young stellar populations.}
\label{fig:sauron}
\end{figure}

\begin{figure}[t]
\centering
\includegraphics[width=3.4in,clip, trim=0cm 2cm 0cm 1.8cm]{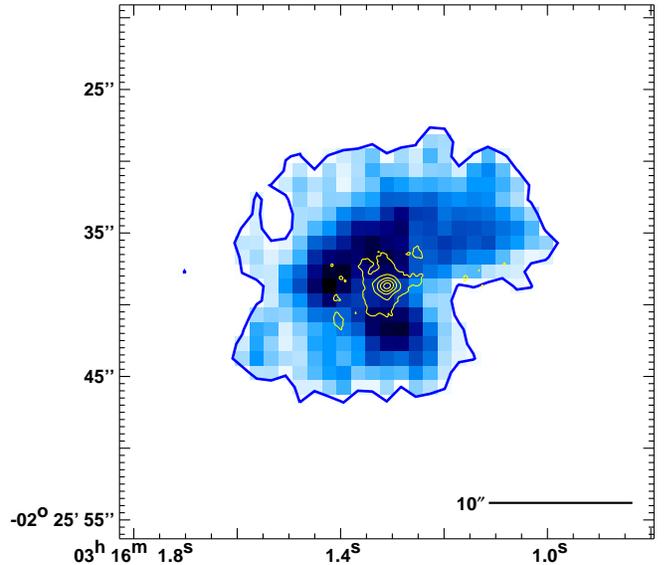}
\caption{ {\em Swift} UVOT {\em UV} image (bluescale), with the $3\sigma$ boundary appearing as a blue contour, overlaid with the CO(1--0) integrated intensity map (moment0; yellow contours).  The {\em UV} image indicates the region which contains young (but not nascent) stars, compared to the current location of the molecular gas.  The total extent of the CO is $\approx 250$ pc, with the most compact molecular gas located only in the central 100 pc \citep{alatalo+11}.  The young stars, as traced by the {\em UV}, are more extended, up to 12\arcsec, or 2 kpc.}
\label{fig:sf_distrib}
\end{figure}

In contrast, young stellar populations, as traced by stellar absorption line strengths (Fig. \ref{fig:sauron}; McDermid et al., in prep), are found throughout the {\tt SAURON} field-of-view ($40'' = 6$ kpc).  A map of the SSP-equivalent ages in NGC~1266 indicates that the youngest stellar populations are concentrated towards the center of the galaxy, particularly within the central 5--6\arcsec (the central kpc; Figure \ref{fig:sauron}). The measured single stellar population ages range from approximately 1.1~Gyr in the central kpc to about 2 Gyr at larger radii, agreeing well with the {\tt pPXF} fit to the \citet{moustakas+06} spectrum discussed in \S\ref{spec_analysis}.  These SSP ages represent a luminosity-weighted combination of young and old stellar populations, so the actual ages of the young stellar populations cannot be directly determined.  The UV emission, known to trace young (as opposed to nascent) stars \citep{kennicutt98} also shows that there are young stars over a larger scale than the molecular gas (see Fig. \ref{fig:sf_distrib}), out to a radius of 2 kpc.

The combined stellar population data and star formation tracers paint an intriguing picture of NGC~1266.  Less than 1~Gyr ago, star formation was occurring on large scales ($\gtrsim 6$~kpc, the edge of the {\tt SAURON} field-of-view).  However, any ongoing or future star formation will be fueled by the available molecular gas and thus only occur in the central 250~pc.  In NGC~1266, we may be observing an object just as it is transitioning into a post-starburst phase, in which widespread star formation has ceased over the course of $\lesssim 1$~Gyr.  Central star formation is expected to be suppressed rapidly ($< 100$ Myr; \citealt{alatalo+11}) in the center of NGC~1266, corresponding with the ignition of the AGN.

\begin{figure*}[t]
\centering
\subfigure{\includegraphics[height=3.8in,clip,trim=0cm 0cm 0cm 1cm]{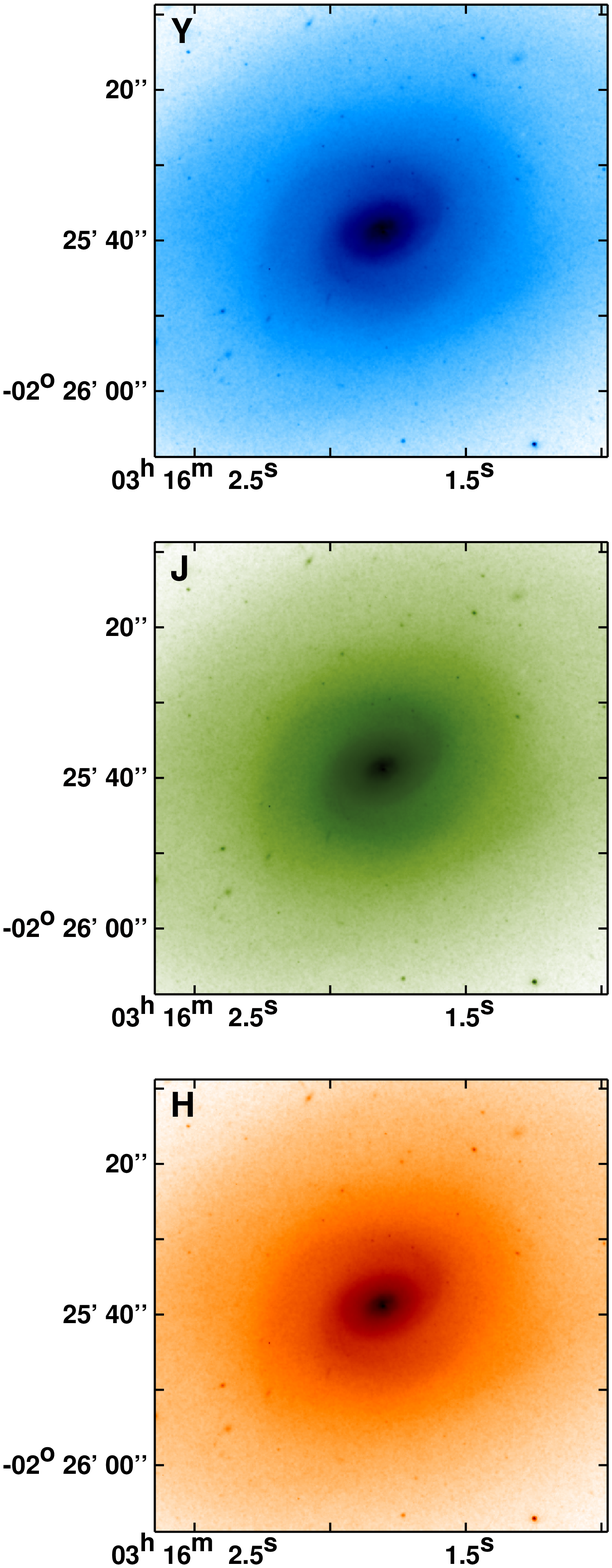}}
\subfigure{\includegraphics[height=3.8in]{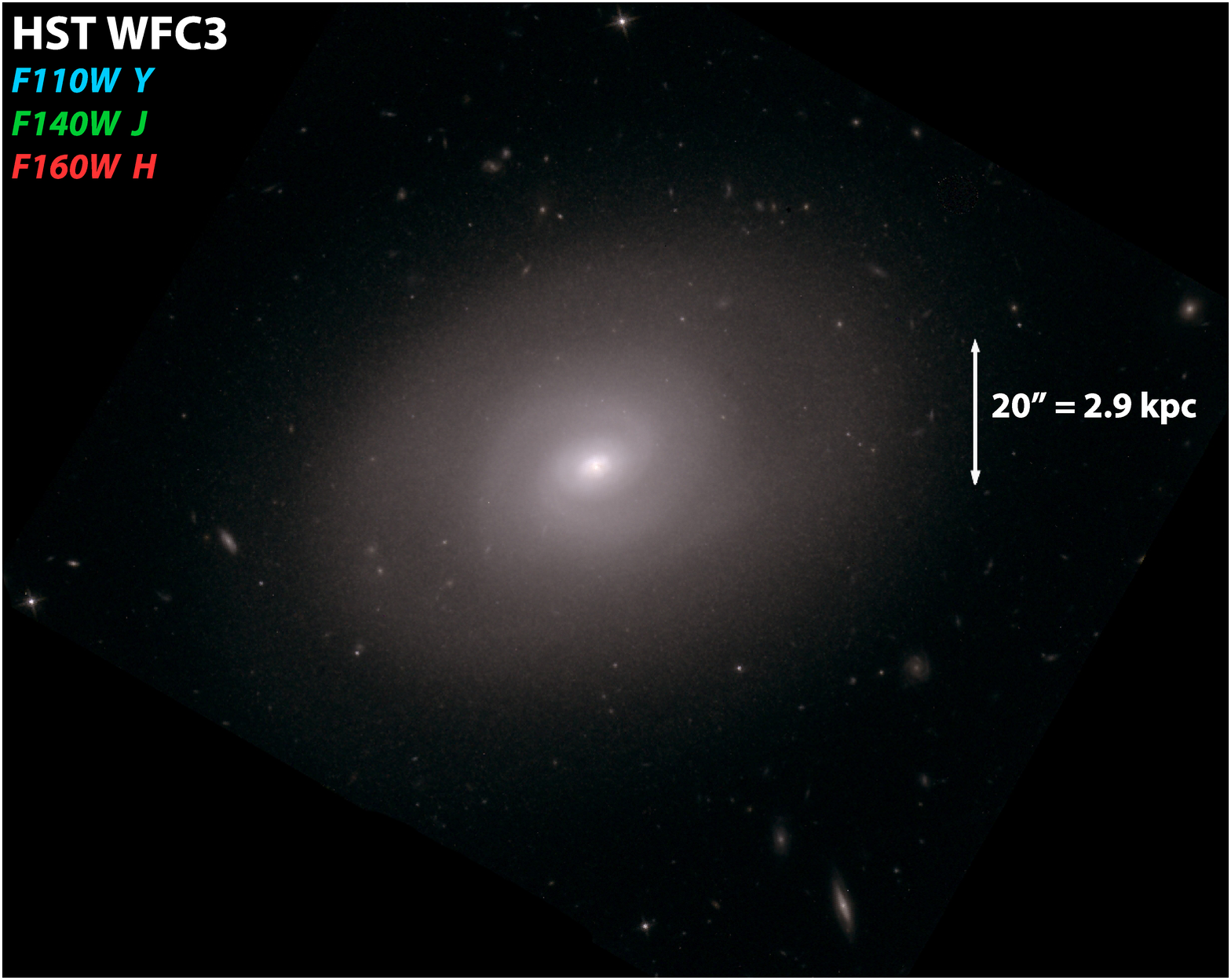}}
\caption{{\bf (Left):} 3-panel figure showing the calibrated emission from HST WFC3.  The bluescale image corresponds to the $Y$ band, the greenscale image corresponds to the $J$ band, and the redscale image corresponds to the $H$ band.  {\bf (Right):} 3-color image constructed from the $Y$ , $J$ and $H$ bands corresponding to B, G and R respectively. The near-IR bands from HST show that there is an underlying spiral structure in the galaxy, previously un-discovered from ground-based observations.  The spiral structure seen here appears to be dynamical in nature, meaning induced by tidal torques, rather than the consumption of gas, which would create a more blue color to the spirals.}
\label{fig:wfc3color}
\end{figure*}

\subsection{Star formation history of NGC~1266}
In order to determine whether NGC~1266 would classically be considered a post-starburst galaxy, we followed the definition of \citet{quintero+04}, that K+A galaxies have A/K light fraction ratios $\gtrsim 0.2$.  The A/K fraction for the nucleus of NGC~1266 is 5.7 (it is 2.1 for the integrated spectrum).  NGC~1266 would thus be classified as a post-starburst, confirmed by the fraction of $< 2$ Gyr stars from the multi-component stellar population fit.
%{\bf Indeed, the multi-stellar population fit indicates that the fraction of $<2$ Gyr stars is even larger, almost half of the total stellar light contribution in the nucleus and about a third of the light in the outer spectrum.}

The post-starburst classification requires both a young stellar population and a reduction of current SF.  The nucleus of NGC~1266 has an A/K stellar ratio that satisfies the post-starburst condition, and the spatial distribution of young stars compared to the molecular gas show that the sites of current SF have changed drastically, with a 2~kpc SF disk in the past to a $>100$pc disk presently.  Although the evidence supports the suggestion that NGC~1266 is transitioning to a post-starburst system, the presence of strong H$\alpha$ emission would likely disqualify NGC~1266 from being classified as a K+A galaxy in a standard post-starburst search, since copious H$\alpha$ emission is typically associated with current SF.  However, \citet{davis+12} showed that the ionized emission in NGC~1266, in particular when investigating the resolved log([O~III]$\lambda5007$/H$\beta$) vs log([S~II]$\lambda{6717,6731}$/H$\alpha$) diagnostic lies in the LINER region of \citet{kewley+06}, and so cannot be primarily from star formation.

The Integral-Field Unit (IFU) spectra from {\tt SAURON} and the GMOS IFU on the Gemini North telescope presented in \citet{davis+12} revealed \ha, H$\beta$, [\ion{O}{3}], [\ion{O}{2}], [\ion{N}{2}] and [\ion{S}{2}] emission within the central 2~kpc of the galaxy.  Spatial maps, line ratios and the high velocities of the ionized gas emission were all most consistent with a shock origin, and \citet{davis+12} argued that the majority of \ha\ emission seen in NGC~1266 is not due to \ion{H}{2} regions from SF but instead arises from shocks.  %Thus, NGC~1266 fits the post-starburst criterion quite well, with both evidence of a young stellar population as well as a lack of evidence for current SF.

Secular processes alone, such as stellar mass-loss, will most likely not be able to significantly replenish the depleted cold gas over a short timescale and ignite any significant level of SF.  This brings to light the possibility that classic searches for post-starburst galaxies are missing some exciting specimens: those in which an AGN is actively expelling molecular gas.  If NGC~1266 is any indication, systematic searches for post-starburst galaxies should not a priori reject all galaxies with strong H$\alpha$ emission.  Instead, such searches should consider ratios of various ionized gas emission lines and search for indications that the emission is due to shocks rather than SF.

Given the typical timescale during which a post-starburst galaxy is detectable as such (1~Gyr; \citealt{quintero+04}), $\sim10$\% of the post-starburst population could be in the process of an NGC~1266-like molecular gas expulsion (assuming a $<100$ Myr timescale for this event; \citealt{alatalo+11}).  A significant fraction of these post-starburst galaxies with outflows would be rejected from standard searches due to the presence of \ha\ emission from shocks.  Because the current state of NGC~1266 represents such an important stage in a galaxy's evolution from actively star-forming to quiescent, it is essential that we modify post-starburst search routines to include NGC~1266-like objects.

\begin{figure*}
\centering
\includegraphics[width=6.5in,clip,trim=0.5cm 4cm 0.5cm 3.8cm]{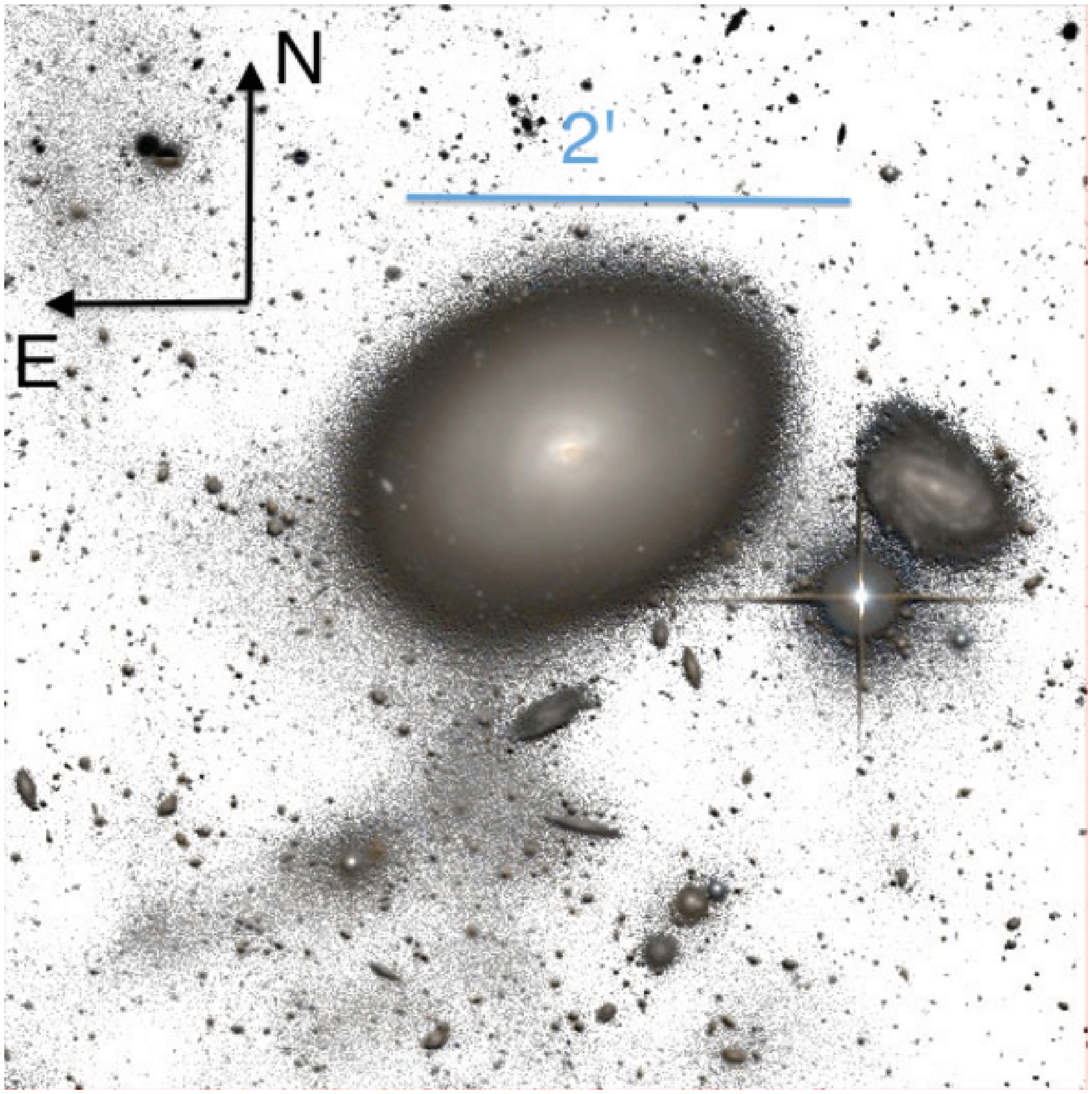}
\caption{$g$ and $r$ wide-field imaging of NGC~1266 taken with the MEGACAM instrument on the Canada-France-Hawaii Telescope on Mauna Kea as part of the MATLAS survey \citep{duc+11,paudel+13}.  The image shown has a limiting surface brightness of 28.5 mag arcsec$^{-2}$.  Field of view is 5\farcm4$\times$5\farcm4 (or $47\times47$ kpc$^{2}$).  There are possibly faint tidal streams seen below and to the left of the galaxy due to material from a disrupted dwarf, but it is clear from these data that there is no sign of a major interaction in the past Gyr.  The large-scale feature seen to the south is likely part of a large Galactic cirrus complex.  Spectroscopy of the spiral galaxy to the east shows it to be in projection in the background (J. Silverman, private communication).}
\label{fig:megacam}
\end{figure*}

\vskip 1cm

\section{Discussion}
\subsection{The connection between the post-starburst and the molecular gas in the center}

Now that it has been established that NGC~1266 hosts a large amount of molecular gas, with a non-negligible fraction of that outflowing molecular gas \citep{alatalo+11}, as well as a post-starburst stellar population, the natural next step is examining the connection between these two properties.  Many authors have suggested a causal link between the AGN and the quenched SF \citep{Hopkins+05,nesvadba+08,cano-diaz+12}, with the AGN and its outflow directly quenching the SF seen in galactic disks.  Although ongoing AGN feedback may be able to disrupt SF in the nucleus of NGC~1266 \citep{alatalo+11,davis+12,nyland+13}, evidence that the current level of AGN feedback is responsible for the decline in SF on the 2 kpc scales of the stellar disk is currently lacking.  \citet{alatalo+11} and \citet{davis+12} demonstrate that the timescale which the AGN has been impacting the gas is about 3~Myr, about two orders of magnitude smaller than the age of the stars in the post-starburst disk.  The short timescale of current AGN activity and post-starburst timescale simply do not agree.
%In NGC~1266, it is clear that the cessation of SF in the 2kpc disk is not causally linked to the {\em current} AGN activity.  
%It is unlikely that the AGN was able to act on the molecular kpc scales, but leave the nuclear gas (which it is much more easily able to influence) largely un-impacted for the past 500 Myr, only to ignite and impact that nuclear gas 3 Myr ago.  

Observations from green valley SDSS galaxies offer another possible explanation.  In SDSS, \citet{schawinski+07} observe that galaxies with AGNs in the green valley appear to have had its most recent SF episode between 100~Myr and 1~Gyr prior, which seems to be what we observe in NGC~1266.  Assuming that the 1~kpc post-starburst disk were just sub-critical at the time of the minor merger, it is possible that the gravitational torques from the dwarf galaxy would be able to trigger a starburst, while simultaneously funneling the molecular gas into the nucleus.  From \citet{cappellari+13}, the total stellar mass within a 1~kpc radius is $\approx 3.9\times10^9$ \msun.  The young mass fraction from the integrated spectral {\tt pPXF} fit, assuming the Kroupa universal IMF is 8\%.  This means that there are a total post-starburst stellar mass of $\approx 3.1\times10^8$ \msun.  The current mass of molecular gas in the system ($1.7\times10^9$ \msun; \citealt{alatalo+11}) is a factor of 5 larger.  The light fraction weighting seen in figures \ref{fig:int_spec} and \ref{fig:nuc_spec} seems to indicate that the young stellar population is older (about 0.8 Gyr) at 1 kpc than at 100 pc, indicating that it is possible that the youngest stars are closest to the nucleus.  This might mean that instead of the post-starburst being driven directly by the AGN, a dynamical event 500~Myr ago was responsible both for the post-starburst population seen today, as well as driving the molecular gas into the nucleus, which is currently being impacted by the AGN.

\subsection{Triggering gas inflow to the nucleus}

The most common trigger invoked when discussing SF quenching is a major merger, but in NGC~1266 this scenario is highly unlikely.  NGC~1266 does not show evidence of a major merger, either kinematically or morphologically.  The stellar kinematics of NGC~1266 were mapped as part of the \atlas\ project \citep{krajnovic+11}, and are undisturbed. Near-IR ({\em Y-J-H}) images of NGC~1266 from the HST do not show any significant morphological disturbance, as one would expect from a major merger, although they do show a weak stellar spiral (Fig. \ref{fig:wfc3color}).  The lack of morphological and kinematic disturbances places a constraint on the possible triggers for the post-starburst, eliminating major mergers (those with a mass ratio of more than $\approx$ 9:1; \citealt{lotz+10}) within the last 1~Gyr.  

On the other hand, mild gravitational encounters are capable of exciting long-lived ($\gtrsim 1$ Gyr) spirals \citep{chang+11,struck+11}, similar to the spiral imprint seen in Fig. \ref{fig:wfc3color}.  \citet{chang+11} show that a perturbing companion as small as 1:100 is capable of creating the spiral pattern seen, though \citet{struck+11} argue that the encounter should be at least a 1:10 mass ratio, which also has the added benefit that it is capable of driving the molecular gas into the center \citep{mihos+94,chang08,struck+11}.  High-sensitivity, wide-field MEGACAM $g$ and $r$ imaging of NGC~1266 are presented in Fig. \ref{fig:megacam}.  The MEGACAM image was taken as part of the MATLAS survey\footnote{http://irfu.cea.fr/Projets/matlas/}. The acquisition and analysis of the MATLAS galaxies is identical to that described for the New Generation Virgo Survey (NGVS; \citealt{ferrarese+12}) described by \citet{paudel+13} for NGC~4216.
Figure \ref{fig:megacam} seems to exhibit a weak tidal stream seen southwest of the galaxy, which could very well be the remnants of a disrupted dwarf galaxy, as well as is able to trace the spiral structure seen in Fig. \ref{fig:wfc3color} to larger radii, being visible at least 10 kpc from the center.  It is therefore possible that the spiral structure seen in Fig. \ref{fig:wfc3color} and the faint tidal tail in Fig. \ref{fig:megacam} could be imprints of a minor merger event that triggered the post-starburst and shepherded the remaining molecular gas into the center of NGC~1266 about 500 Myr ago.

If both the post-starburst and the nuclear gas are indeed connected by a common trigger, then we must explain why the molecular gas still remains in the nucleus, when one event occured in the past 1/2~Gyr and the AGN outflow has only existed for the past 2.6~Myr \citep{alatalo+11}. One can imagine that there might be a lag time between the ignition of the AGN and the funneling of gas into the nucleus, with many suggested mechanisms able to stall the AGN.  One hypothesis is that the transport of gas from large radii forces the gas to assume some distribution in angular momentum, i.e., the stochastic model of AGN fueling \citep{King2007, Nayakshin2007}.  Another possibility is that cloud-cloud collisions are enhanced at small radii as initially suggested by \citet{Shlosman1990}.  A radiation pressure supported disk has also been suggested by \citet{Thompson2005} (see the discussion in \citealt{chang08}).  Another option is bottlenecking: competition of fuel between SF and BH, whose mechanism for fueling at the pc-level scales is likely internal instabilities \citep{alexander+hickox12}.  Despite the reconciliability of the AGN ignition time and the post-starburst event, the fact that the molecular gas has survived 500 Myr without being completely turned into stars requires further thought.

\subsection{Sustainability of the nuclear molecular gas}

Assuming the ``normal'' star formation efficiency calculated in \citet{leroy+08} of $5.25\times10^{-10}~{\rm yr}^{-1}$, we would expect NGC~1266 to form at least $10^9$ \msun\ worth of stars within 500 Myr, thus exhausting most of the remaining H$_2$ fuel.  Given that the surface density of the nuclear gas ($2.7\times10^4$ \msun\ pc$^{-2}$; \citealt{alatalo+11}) is larger than in a normal galaxy by over two orders of magnitude, we in fact would expect that NGC~1266 would have exhausted its nuclear fuel even more efficiently if it had been sitting at this density within 100~pc of the nucleus for the past 500 Myr.  Stalling SF in the hypercompact molecular disk would therefore require an injection of energy designed to keep the disk stable against gravitational collapse to explain the gas disk's sustained existence.  There seems to be evidence of a population of just such sources, with largely suppressed SF rates: radio galaxies \citep{okuda+05,ogle+07,papadopoulos+08,ogle+10,nesvadba+10,guillard+12}.  \citet{nesvadba+10} show that radio galaxies as a group are inefficient starformers, with efficiencies 10--50 times smaller than in normal galaxies.  They go on to hypothesize that the turbulent kinetic energy injection from the radio jet is able to stall star formation in the molecular gas for $\approx 10^8$ yr (as compared to the radio jet time, of $\approx 10^7$ yr).  This order-of-magnitude difference in timescales is explained through a cascading deposition of turbulent energy from large scales to small scales, originally suggested for the shocked region of Stephan's Quintet by \citet{guillard+09}.  This would allow the molecular gas in the nucleus to remain SF inefficient for $\gtrsim100$ Myr timescales, if ignition and momentum transfer of the radio jet occurs periodically.

Recent Very Long Baseline Array (VLBA) findings of \citet{nyland+13} show that NGC~1266 has a compact radio source, attributed most likely to the AGN.  The radio lobes originally described in \citet{baan+06} are suggested as being part of a radio jet in \citet{alatalo+11}, \citet{davis+12} and \citet{nyland+13}.  This could mean that NGC~1266 is a radio galaxy whose jet just turned on (about 3 Myr ago; \citealt{alatalo+11}).  \citet{schoenmakers+00} have shown that many radio jets in galaxies are cyclical phenomena, fading about $\sim10^7$ yr after the initial jet creation.  This would mean that the radio jet is able to impact the molecular gas $\sim10$ times longer than it is visible.  It is therefore possible that the radio jet in NGC~1266 could be a cyclical phenomenon, leaving an impact on the SF in molecular disk without the outward signs of radio synchrotron lobes (which have already faded).  In fact, a 365 MHz radio survey conducted by \citet{douglas+96} indicates that NGC~1266 appears to have be double lobed structure, 25\arcsec\ (or $\approx$ 4 kpc) in size, at least twice as large as the 1.4 GHz-detected radio jets \citep{alatalo+11,nyland+13}.  This conclusion was reached by fitting the interferometric data to a model, therefore it is important to confirm the existence of the low frequency double-lobed jet through radio imaging.

If the radio jet feedback seen in NGC~1266 is indeed cyclical, then gas is currently being heated and deposited into the galactic halo, which is commonly seen in galaxy clusters and groups (\citealt{gitti+12}, and references therein).  In most cases where X-ray bubbles are seen, the gas is being deposited into the media of either a group or a cluster.  NGC~1266 appears to be isolated \citep{cappellari+11b}, and therefore does not have access to the combined gravitational potential of many nearby neighbors.  \citet{boroson+11} do confirm that hot gas haloes are present around isolated (and less massive) galaxies, but at a much lower luminosity.  This is likely because much of the gas is able to escape the gravitational potential of the galaxy, and therefore would be unavailable for future recycling.  One conclusive way to confirm whether or not the SF quenching episode is due to this cyclical AGN feedback mechanism is with deep X-ray and low frequency radio observations to search for fossil shells from previous episodes, noting the 365 MHz interferometric data provide compelling evidence that this may be the case for NGC~1266.

It is also possible that these two events: the 500 Myr old post-starburst and the current expulsion of the molecular gas, are not related at all.  If this is the case, there are many scenarios that are able to explain the post-starburst event, including the previously discussed gravitational encounter, or possibly an AGN event similar to what is hypothesized in \citet{cano-diaz+12}, that the AGN in the system is directly creating the post-starburst.  The nuclear molecular gas, on the other hand, remains very difficult to explain, as a minor merger capable of depositing all $1.7\times10^9$ \msun\ of gas into the center would have to be at least a 5:1 ratio, and have deposited its gas less than a couple dynamical times ago, which should mean leaving sufficient evidence behind.  Instead, there is little evidence of this, both from the lack of H~I emission \citep{alatalo+11,nyland+13}) as well as the lack of prominent tidal features (Fig. \ref{fig:megacam}).  

Given the observational evidence, it seems the current favored explanation for the current state of molecular gas is that a gravitational encounter 500 Myr ago was able to drive the molecular gas into the nucleus, prompting first a starburst, then quenching star formation within the 1~kpc disk.  The spiral structure seen in Figs. \ref{fig:wfc3color} and \ref{fig:megacam} supports this picture.  The torques also caused the molecular gas to funnel into the nucleus, where the AGN was able to inhibit its ability to form stars, through the injection of turbulence from the radio jets.

%%%%%%%%%%%%%%%%% CONCLUSION %%%%%%%%%%%%%%%%%%%%%	
\section{Summary and Conclusions}
\label{conclu}%% start new page for next chapter
We report new observations of NGC~1266, a local example of an AGN-driven molecular outflow and candidate for AGN-driven SF quenching.  

An investigation of the spatial distribution of young stars using the {\em Swift} UVOT {\em UV} band as well as {\tt SAURON} stellar absorption and age maps, compared with the molecular gas distribution, shows that the sites of current star formation are far more compact than the regions containing young stars.  This points to an outward-in cessation of star formation within the galaxy and suggests that NGC~1266 might be transitioning into a post-starburst object.

We also performed a stellar population analysis of NGC~1266 utilizing the spectra from \citet{moustakas+06}.  The absorption features in the nucleus of NGC~1266 indicate the presence of a non-negligible fraction of $<1$ Gyr-aged stars.  The model-derived A/K fraction of NGC~1266 of 2.1 would lead it to be classified as post-starburst in SDSS, but previous studies have likely failed to recognize the post-starburst nature of NGC~1266 due to the presence of strong ionized gas emission.  However, as \citet{davis+12} demonstrated, the ionized gas in NGC~1266 is most likely the result of shocks associated with the outflow rather than SF.  NGC~1266-like post-starbursts may be rejected by standard post-starburst searches due to the presence of the ionized gas emission, and it is therefore imperative to expand the search for post-starburst candidates to include galaxies with shock-like line ratios.

The post-starburst stellar population within NGC~1266 sets a timescale for the quenching event of $\approx 500$ Myr ago, which seems to indicate that current AGN activity cannot be directly responsible for the cessation of SF in the 2-kpc disk.  Instead, it is possible that a gravitational encounter, which excited spirals within the galaxy, was able to drive the molecular gas into the center, thus quenching any possible SF outside of $\approx 100$pc of the nucleus.  Recent studies of radio galaxies seem to indicate that the AGN then might be able to prevent the nuclear molecular gas from forming stars by injecting turbulent kinetic energy via a periodical ignition of the radio jet.

%No prominent tidal features have been seen in NGC~1266 even with deep optical imaging, indicating that the mechanism responsible for quenching star formation cannot be a major merger, and in fact, the timescales seem to support the hypothesis that a minor merger was responsible for the event.  This leads to the possibility that SF quenching in some systems is an induced-quenched process, beginning with an agitating minor merger to transport the gas into the center and ending when the AGN is triggered and drives the nuclear gas completely out of the system.

\vskip 1cm
%%%%%%%%%%%%\section*{Acknowledgments}
\acknowledgments{{\it Acknowledgments:}  KA would like to thank Carl Heiles, Phil Appleton and Joan Wrobel for many useful discussions and input, as well as John Moustakas for access to the most up-to-date version of the long-slit spectra.  KA would also like to thank the anonymous referee for prompt replies and insightful suggestions that have vastly improved the paper.  The research of KA is supported by the NSF grant AST-0838258 and by NASA grant HST-GO-12526. KA was also partially supported by funding through Herschel, a European Space Agency Cornerstone Mission with significant participation by NASA, through an award issued by JPL/Caltech.  KN is supported by NSF grant 1109803.  SLC was supported by ALMA-CONICYT program 31110020.  PS is an NWO/Veni fellow. RMMcD is supported by the Gemini Observatory, which is operated by the Association of Universities for Research in Astronomy, Inc., on behalf of the international Gemini partnership of Argentina, Australia, Brazil, Canada, Chile, the United Kingdom and the United States of America. RLD and MB are supported by the rolling grants ÔAstrophysics at OxfordÕ PP/E001114/1 and ST/H002456/1 from the UK Research Councils. RLD acknowledges travel and computer grants from Christ Church, Oxford and support from the Royal Society in the form of a Wolfson Merit Award 502011.K502/jd. TN acknowledges support from the DFG Cluster of Excellence: ÔOrigin and Structure of the UniverseÕ.  Based on observations made with the NASA/ESA Hubble Space Telescope, obtained from the Space Telescope Science Institute, which is operated by the Association of Universities for Research in Astronomy, Inc., under NASA contract NAS 5-26555. These observations are associated with program \#12526.  This work has made use of the tools and services available through the HEASARC online service, which is operated by the Goddard Space Flight Center for NASA.  We acknowledge the use of public data from the Swift data archive.
\\
%%%%%%%%%%%%%%%%%%%%%%%%%%%%%%%%%%%%%%%%%%%%%%%

% BIBLIO
\bibliographystyle{apj}
%\bibliography{ms}

 %%%%%%%%%%%%%%%%%%%%%%%%%%%%%%%%%%%%%%%%%%%%%%%
% END

 \end{document}